\pdfoutput=1
\documentclass[12pt]{iopart}
\usepackage{iopams}
\usepackage{graphicx}
\usepackage{color}

\begin{document}
\title{Equilibrium winding angle of a polymer around a bar}
\author{J.-C. Walter}
\address{Institute for Theoretical Physics, KULeuven, 
Celestijnenlaan 200D, B-3001 Leuven, Belgium}
\author{G. T. Barkema}
\address{Institute for Theoretical Physics, Utrecht University, 
Leuvenlaan 4, 3584 CE Utrecht, The Netherlands}
\address{Instituut-Lorentz for Theoretical Physics, University of Leiden, 
Niels Bohrweg 2, 2333 CA Leiden, The Netherlands}
\author{E. Carlon}
\address{Institute for Theoretical Physics, KULeuven, 
Celestijnenlaan 200D, B-3001 Leuven, Belgium}

\begin{abstract}
The winding angle probability distribution of a planar self-avoiding walk has been known exactly
since a long time: it has a gaussian shape with a variance growing as $\langle \theta^2\rangle \sim \ln L$. 
For the three-dimensional case of a walk winding around a bar, the same scaling is suggested, based
 on a first-order epsilon-expansion. We tested this three-dimensional case
by means of Monte Carlo simulations up to length $L\approx25\,000$ and using exact enumeration data
 for sizes $L\le20$. We find that the variance of the winding angle scales as  
$\langle \theta^2\rangle \sim \left( \ln L\right)^{2\alpha}$, with $\alpha=0.75(1)$. 
The ratio $\gamma = \langle\theta^4\rangle/\langle\theta^2\rangle^2=3.74(5)$ is incompatible
 with the gaussian value $\gamma =3$, but consistent with the observation that the tail of the
 probability distribution function $p(\theta)$ is found to decrease slower than a gaussian function.
 These findings are at odds with the existing first-order $\varepsilon$-expansion results.
\end{abstract}

\date{\today}

\maketitle

\section{Introduction}

In view of the great importance of DNA in biology and
biotechnology, there has been a considerable interest in the past
years in the modeling of dynamics and equilibrium properties of DNA
molecules~\cite{peyr89,barb99,kafr00,carl02,peyr06,cocc99,rudn02,barb03,kaba09,baie10a}.
From the perspective of statistical physics, DNA is an interesting system
as it shows a broad range of phenomena such as phase transitions in response
to temperature changes and to the application of mechanical forces. A
paradigmatic example is that of the melting transition, i.e. the
separation of the two strands of the helix induced by a temperature
change, which has been the object of considerable attention by the
scientific community~\cite{peyr89,barb99,kafr00,carl02,peyr06}.  Also the
melting {\em dynamics}, and in particular the openings of bubbles, has
been the topic of various studies~\cite{mare01,kunz07,baie09}.  A new
aspect which shows up prominently in the dynamics, while being of less
importance in equilibrium, is the helical nature of the double-stranded
DNA, which lead to models of DNA molecules which approximately include
helical degrees of freedom~\cite{cocc99,rudn02,barb03,kaba09}. Also,
in a recent simulation study involving two of us~\cite{baie10a}, the
scaling relation between the unwinding time $\tau_u$ of a double-helical
structure, as a function of its length $L$, has been studied, with the
result $\tau_u \sim L^{2.58\pm 0.02}$. A theoretical understanding of
this scaling is still missing.

A system closely related to the unwinding of a double-helical structure,
is the unwinding of a single polymer initially wound around a fixed bar.
The advantage of this latter system is that it has a cleanly defined
reaction coordinate: the winding angle of the free end.  While it is
our ultimate goal to understand fully the dynamics of unwinding, the
topic of the current research is restricted to identify the equilibrium
properties of a single polymer wound around a bar, thereby providing a
solid basis for further research.

Further motivation behind our study, is that there has been quite some
recent interest in the study of topological properties of long flexible
polymers, inspired by biology, in the context of chromosomal segregation
(see e.g. Ref.~\cite{mark09}). The scaling behavior of the probability
distribution function of linking numbers of two closed polymer rings
reveals some interesting properties which could explain segregation of
chromatide domains~\cite{mark09}. The winding angle distribution of a
polymer attached to a bar is an analog of the linking number, so it is
natural to ask whether the results discussed here could be extended to
other topological invariants for closed curves.

\section{Winding angle distributions}

The study of winding properties of random and self-avoiding
walks has a long history which dates back to more than
half a century ago~\cite{spit58}. These properties are relevant for
many applications in different domains of statistical physics
\cite{fish84,rudn88,dupl88c,houc92,dros96,prel98}.  In a few simple
cases the probability distribution function (pdf) of winding angles
can be computed exactly.  For instance, the pdf of a planar random walk
winding around a circle of finite radius is~\cite{rudn87}
\begin{equation}
\lim_{L\to\infty}p\left(x=\frac{2\theta}{\ln L}\right) = 
\frac{\pi}{4} \frac{1}{\cosh^2(\pi x/2)},
\label{disk_abs}
\end{equation}
where $\theta$ is the winding angle and $L$ the length of the walk.
These quantities appear in the pdf as a single scaling variable 
$x = 2\theta/ \ln L$.

Planar self-avoiding walks were studied first by Fisher
{\it et al.} \cite{fish84}. On the basis of scaling arguments and
numerical results they suggested that the distribution of winding angles
is gaussian with a variance growing as $\langle \theta^2 \rangle \sim \ln L$
(as opposed to that of planar random walks in which $\langle \theta^2
\rangle \sim (\ln L)^2$, see Eq.~(\ref{disk_abs})).  These findings were
corroborated by an exact distribution \cite{dupl88c}
\begin{equation}
\lim_{L\to\infty}p\left(x=\frac{\theta}{2\sqrt{\ln L}}\right) = 
\frac{1}{\sqrt\pi}  e^{-x^2},
\label{saw_2d}
\end{equation}
obtained some years later by conformal invariance techniques.

The winding angle distribution of several other polymer systems
were considered, as for instance two-dimensional polymers with
orientation-dependent interactions~\cite{prel98}. The phase diagram of
these polymers contains different phases, e.g. spiral collapsed,
normal collapsed and swollen, separated by a theta point.  The winding angle
distribution turns out to be universal~\cite{prel98}. It is gaussian with
a variance $\langle \theta^2 \rangle = C \ln L$, where the coefficient
$C$ takes different values in different phases.  The gaussian behavior
is therefore a robust feature for self-avoiding walks in two dimensions.

\begin{figure}[t]
\includegraphics[width=0.75\columnwidth]{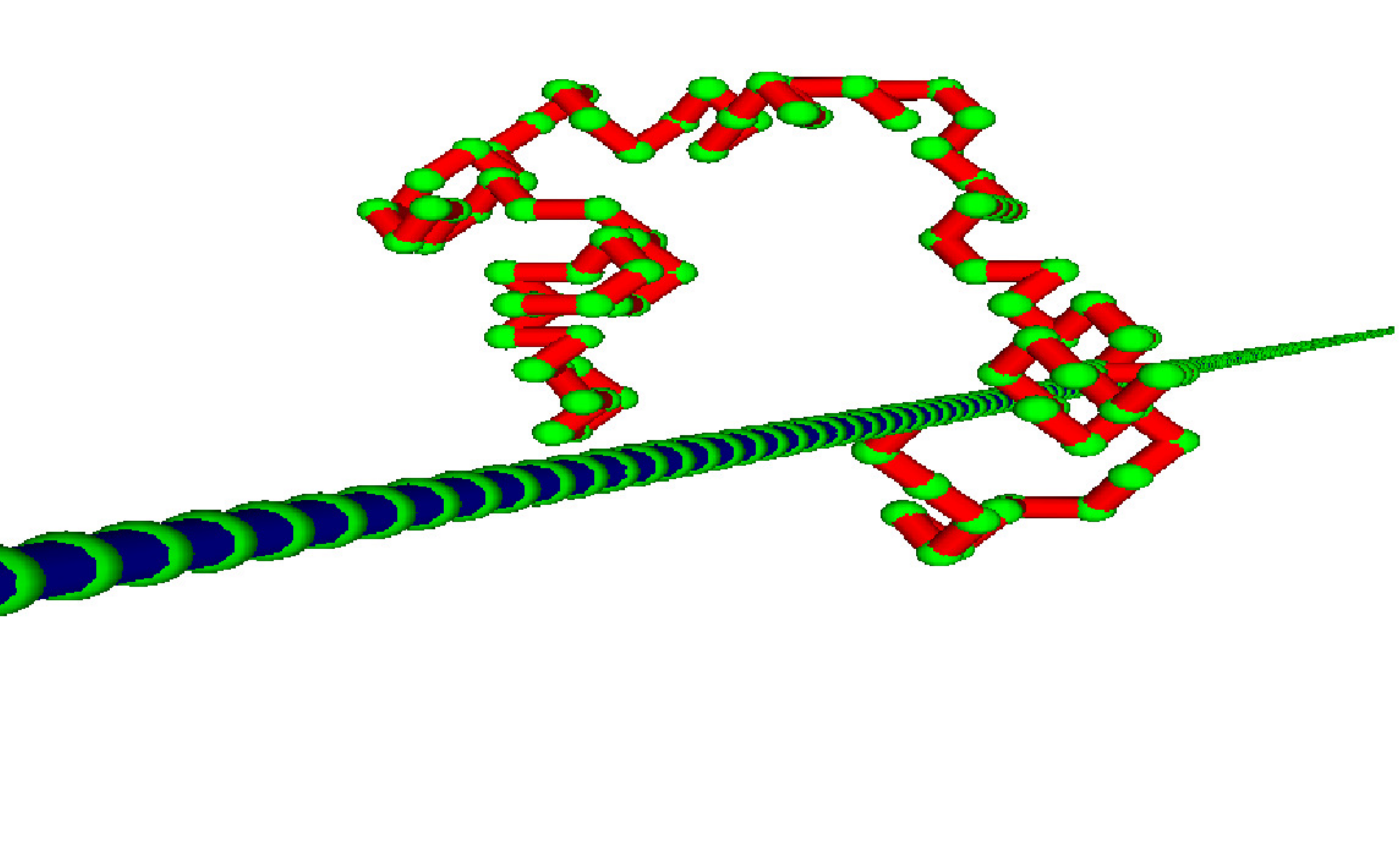}
\caption{Snapshot of an equilibrium Monte Carlo simulation of a 
three-dimensional polymer attached to an infinitely long bar.
The polymer is self-avoiding and has excluded-volume interactions
with the bar.}
\label{SAW3d}
\end{figure}

No exact distribution is known in the case of a polymer attached
by one end to a rigid bar, which is the case studied in this paper
(see Fig.~\ref{SAW3d}). Using renormalization group techniques, in
1988, Rudnick and Hu \cite{rudn88} considered a self-avoiding walk in
$4-\varepsilon$ dimensions whose path wraps around a surface of dimension $2-\varepsilon$.
  The probability distribution for a winding angle $\theta$ of
a polymer of length $L$, at first order in $\varepsilon$, is \cite{rudn88}
\begin{equation}
p(\theta,L,\epsilon) \propto \exp\left( -\theta^2\varepsilon/8 \ln L \right),
\label{sawepsilon}
\end{equation}
which surprisingly matches the exact form of a planar self-avoiding
walk, given in Eq.~(\ref{saw_2d}), by simply setting $\varepsilon=2$.
The authors of Ref.~\cite{rudn88} also performed numerical calculations
of the variance $\langle \theta^2 \rangle$ for self-avoiding walks of
length $L \leq 800$.  Due to the limited computing power available at
that time, only a pre-asymptotic regime could be investigated, which
showed a scaling consistent with that of random walks: $\langle \theta^2
\rangle \sim (\ln L)^2$.

Another interesting class of systems is three-dimensional directed
walks winding around a line~\cite{dros96}, which model the behavior
of flux lines in high-$T_c$ superconductors.  When projected on
the plane perpendicular to the line, the winding problem reduces
to that of a two-dimensional random walk winding around a center.
The winding angle distribution follows again an exponential decay (as
for Eq.~(\ref{disk_abs})), with a decay constant depending on the type of
boundary conditions~\cite{dros96}. The distribution is however a gaussian
in the presence of random impurities in the bulk~\cite{dros96} with the
same scaling variable as in Eq.~(\ref{saw_2d}).  The pinning effect of
the impurities makes the walks meander in the direction away
from the line, with an excursion size growing as $\sim l^{0.62}$ for a
path of length $l$. This behavior is close to that of three-dimensional
self-avoiding walks ($\sim l^\nu$ with $\nu \approx 0.588$), which
may suggest that the winding angle distribution of self-avoiding walks
around a bar would be gaussian. This would also be in agreement with
the $\varepsilon$-expansion distribution (Eq.~(\ref{sawepsilon})), when
setting $\varepsilon=1$. We will, however, show that the winding angle
distribution for a self-avoiding walk around a bar does not follow
gaussian behavior.  In addition the variance $\langle \theta^2 \rangle$
turns out to scale with a non-trivial power of $\ln L$.

\section{The Model}

The numerical results presented in this paper are obtained by Monte
Carlo simulations of lattice polymers up to $L\approx25\,000$ on a
face-centered-cubic lattice and by exact enumerations of short walks ($L
\leq 20$) on a cubic lattice.  In both cases one end of the polymer is
attached to the bar, while the other is free.  In the exact enumeration
study all the possible configurations of a polymer attached at one end
to a bar on a simple cubic lattice were generated using a backtracking
method, and the averaged squared winding angle was computed. 

In the Monte Carlo simulations we performed equilibrium sampling using
two types of updates: (1) local moves, such as corner flips and end-monomer
flips and (2) global moves, involving the rotation of a whole branch of
the polymer from a selected point via the pivot algorithm
 (see~\cite{madr88} for a detailed study of this algorithm).
This algorithm is very efficient because of the small autocorrelation
time. Recently it was applied to the computation of
the growth exponent $\nu$ for very long polymers~\cite{clis10}.  One monomer
is chosen randomly as the pivot point and a random operation of symmetry
allowed by the lattice (rotation or reflection) is applied to the branch
of the polymer not attached to the bar. This attempt is accepted if it
satisfies the self-avoidance.

\begin{figure}[t]
\includegraphics[width=0.9\columnwidth]{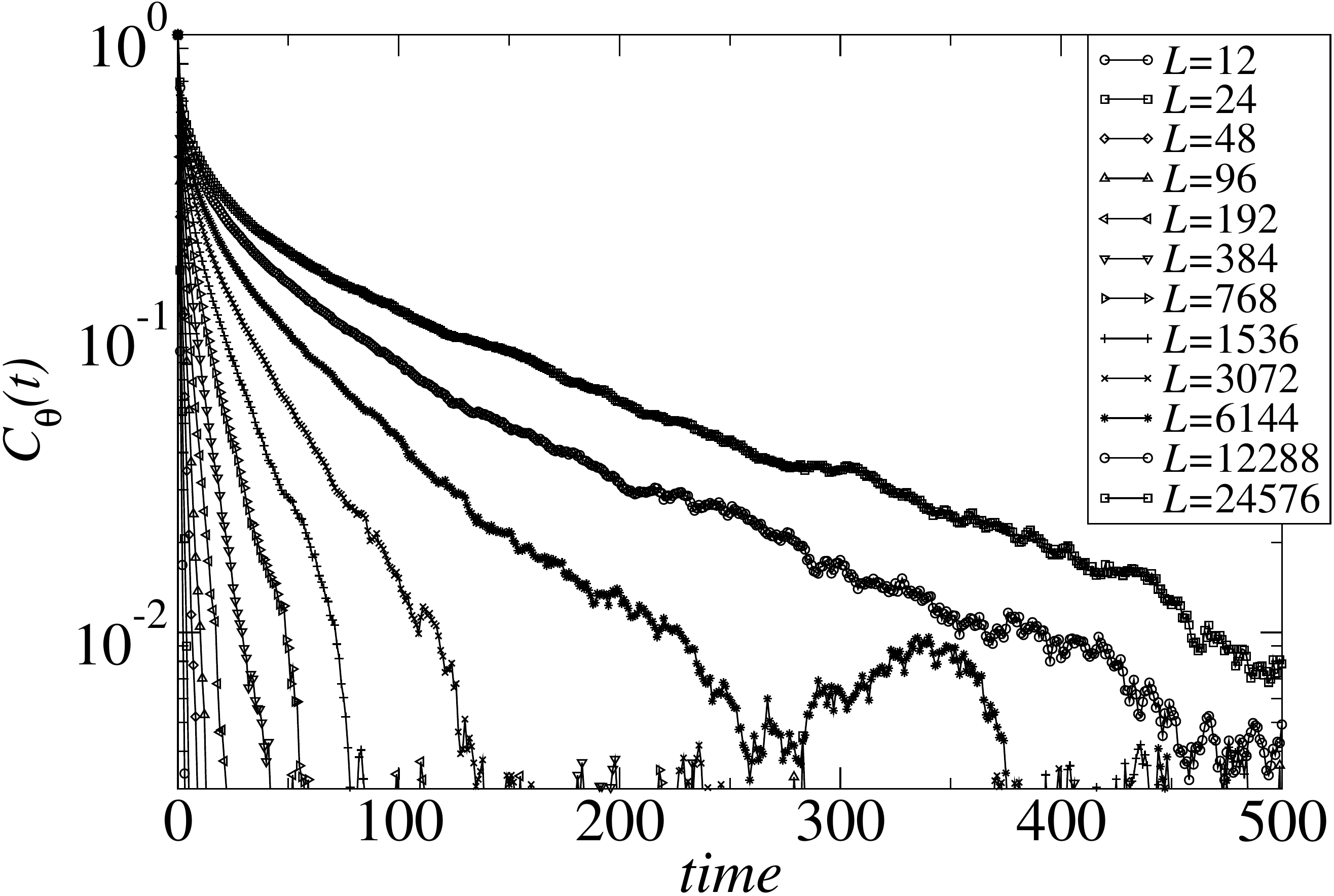}
\caption{Autocorrelation function of the winding angle calculated for
sizes $L=6\cdot2^n$ with $n\in[1;8]$ (from bottom to top). A unit
of time corresponds to $L$ attempts with corner flips and 15 attempted pivots.
The correlation time is estimated through the integration of these
curves~\cite{newm99}.}
\label{corre_theta}
\end{figure} 

The acceptance ratio $R$ (the ratio of accepted moves over the total
number of attempts) of the pivot algorithm scales as 
$R\propto L^{-p}$ where $p\approx0.1918(13)$ in two dimensions and
$p\approx0.1069(9)$ in three dimensions~\cite{madr88}.  Our estimates for $R$ compare well
with those of~\cite{madr88}, but are however slightly smaller due to
collisions with the bar. For the range of sizes investigated in this article,
the acceptance ratio varies between $R \approx0.6$ ($L\approx100$)
and $R\approx0.33$ ($L\approx25\,000$).  In the following, one unit of
time consists of 15 attempted pivots moves, on average corresponding
to 5 accepted pivot moves for the biggest sizes, and $L$ attempted
corner flips.  Since the structure of the FCC lattice allows a winding over
an angle of $2\pi$ within 6 links, the maximum winding angle is $2\pi L/6$. We
choose the polymer length as a multiple of 6: $L=6\cdot2^n$ with $2
\leq n \leq 8$ i.e. up to $L=24576$.  The correlation time $\tau$ is
estimated through the autocorrelation function $C_{\theta}(t)$ of the
winding angle $\theta$. It is defined as
\begin{equation}
C_{\theta}(t)=\frac{\langle\theta(0)\theta(t)\rangle-\langle\theta\rangle^2}
{\langle\theta^2\rangle-\langle\theta\rangle^2},
\end{equation}
where the symbol $\langle\cdots\rangle$ indicates the equilibrium
average.  The results are shown in Figure~\ref{corre_theta}. From the
data we estimate that $\tau \approx 80$ for $L=24576$. We performed a
thermalization during at least $20\tau$, followed by samplings separated
by at least $2\tau$.  We considered polymers of lengths from $L=12$ to
$L=24576$, each time separated by a factor 2. Since the samplings can be
 considered as independent, the fluctuations are estimated using the central 
limit theorem~\cite{newm99}.

\begin{figure}[t]
\includegraphics[width=0.9\columnwidth]{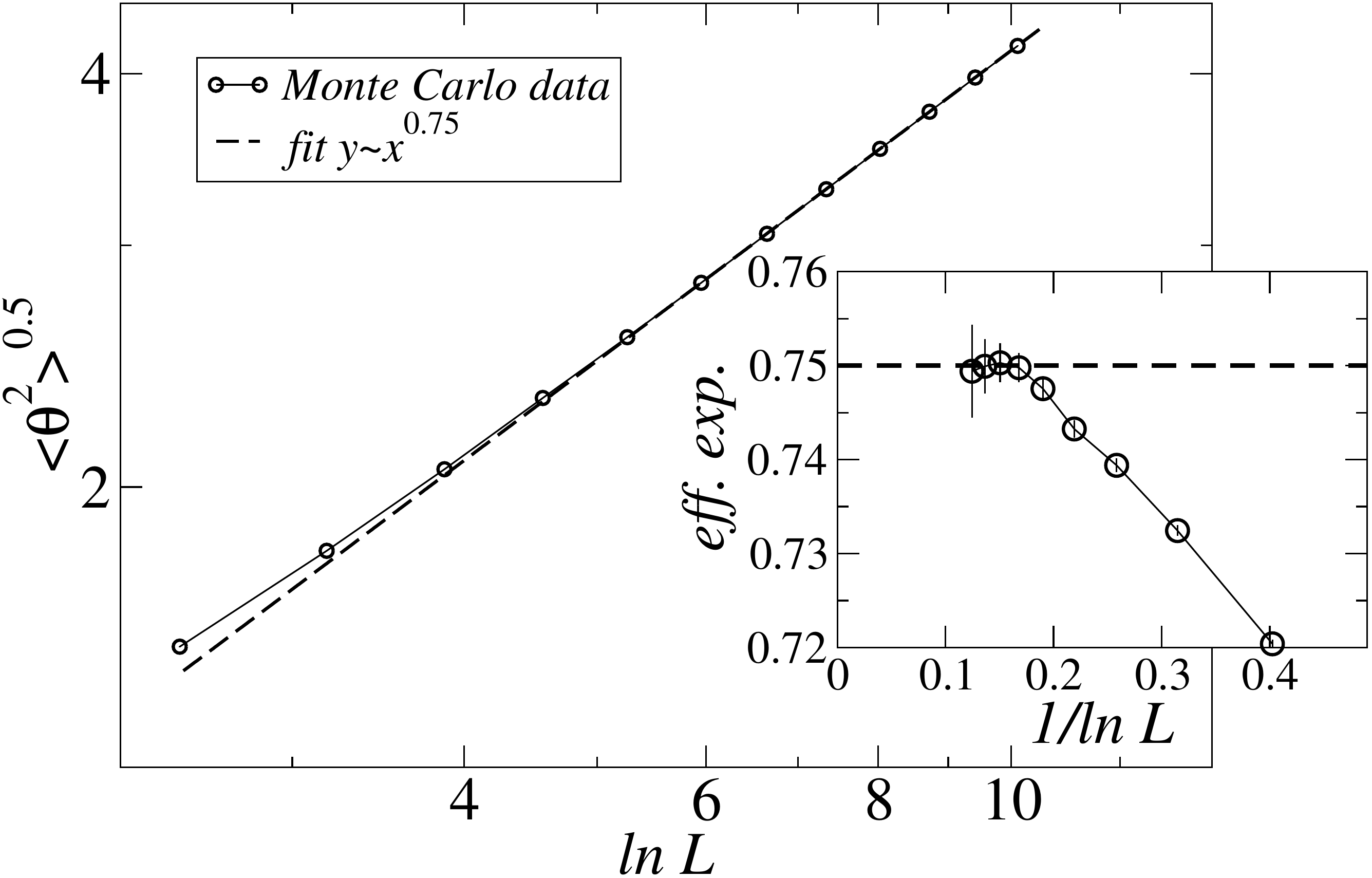}
\caption{Double-logarithmic plot of $\sqrt{\langle \theta^2\rangle}$
vs. $\ln L$ for $L=12$ up to $24576$ with steps of factors of
$2$. The data for $L > 400$ are described within error bars by a
power-law $\sqrt{\langle \theta^2\rangle}\propto(\ln L)^{\alpha}$ with
$\alpha=0.75(1)$. Inset: effective exponent as a function of $1/\ln L$.}
\label{theta2_vs_L}
\end{figure}

\section{Results}

\subsection{Analysis of $\langle \theta^2 \rangle$}

Figure \ref{theta2_vs_L} shows a double-logarithmic plot of $\sqrt{\langle
\theta^2 \rangle}$ as a function of $\ln L$, calculated up to a size
$L_{\rm max}=24\,576$.  Averages were performed over $2\cdot10^6$ and $10^6$
 independent configurations, respectively, for sizes ranging between $L=12$
 and $1536$, and between $L=3072$ and $24\,576$.  
The dashed line represents the best fit to the
data, which implies a scaling of the type $\sqrt{\langle \theta^2
\rangle} \sim (\ln L)^\alpha$ with $\alpha\approx0.75$.  The inset in
Fig.~\ref{theta2_vs_L} shows the effective exponent which was computed
from the slope of the data in the interval $[L, L_{\rm max}]$ for
increasing $L$.  The effective exponent is plotted as a function of
$1/\ln L$ and shows a convergence to $\alpha=0.75(1)$.

\begin{figure}[t]
\includegraphics[width=0.9\columnwidth]{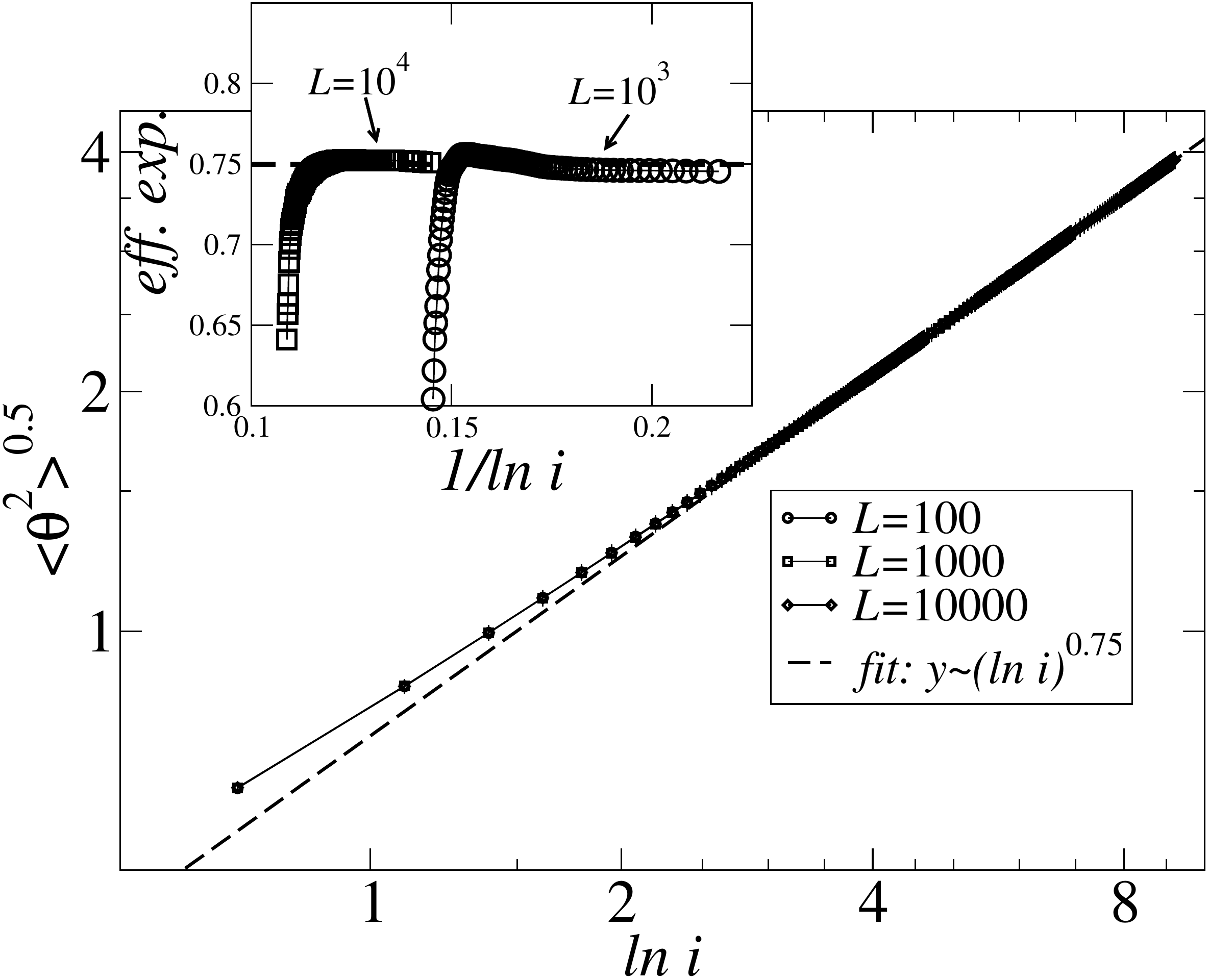}
\caption{Plot of $\sqrt{\langle\theta^2\rangle}$ vs. $\ln i$ (the monomer
index) for three polymers of lengths $L=10^2$, $10^3$ and $10^4$.  Inset:
the effective exponent for the two largest sizes. Disregarding the strong
devations close to the end monomer $i=L$, we estimate $\alpha=0.75(2)$
(dashed line).}
\label{theta2_vs_i}
\end{figure}

A second analysis was performed on polymers with lengths $L=10^2$,
$10^3$ and $10^4$, each averaged over $10^6$ independent configurations.
We computed the winding angle as a function of the monomer index $i$,
which is shown in Fig.~\ref{theta2_vs_i}.  This analysis is done on a
large number of data points as $1 \leq i \leq L$ and yields $\sqrt{\langle
\theta^2 \rangle} \sim (\ln i)^\alpha$ with $\alpha\approx0.75$, again
consistent with the previous estimate 
($i=1$ is the monomer attached to the bar, while $i=L$ is the free end
monomer).  The effective exponent, calculated as above and plotted as a
function of $1/\ln i$, is shown in the inset of Fig.~\ref{theta2_vs_i}.
Starting from small $i$ (right side of the graph), the data quickly
reach a rather constant value, while the effective exponent decreases
for $i \to L$, due to an abrupt change of slope close to the free end
of the polymer.  This behavior is due to end effects and was discarded
from the analysis. We estimated the exponent from the constant plateau
value (dashed line), yielding $\alpha=0.75(2)$. Note that for the
longest polymer $L=10^4$ the region with a constant $\alpha_{\rm eff}$
is considerably broader.

\begin{figure}[t]
\includegraphics[width=0.9\columnwidth]{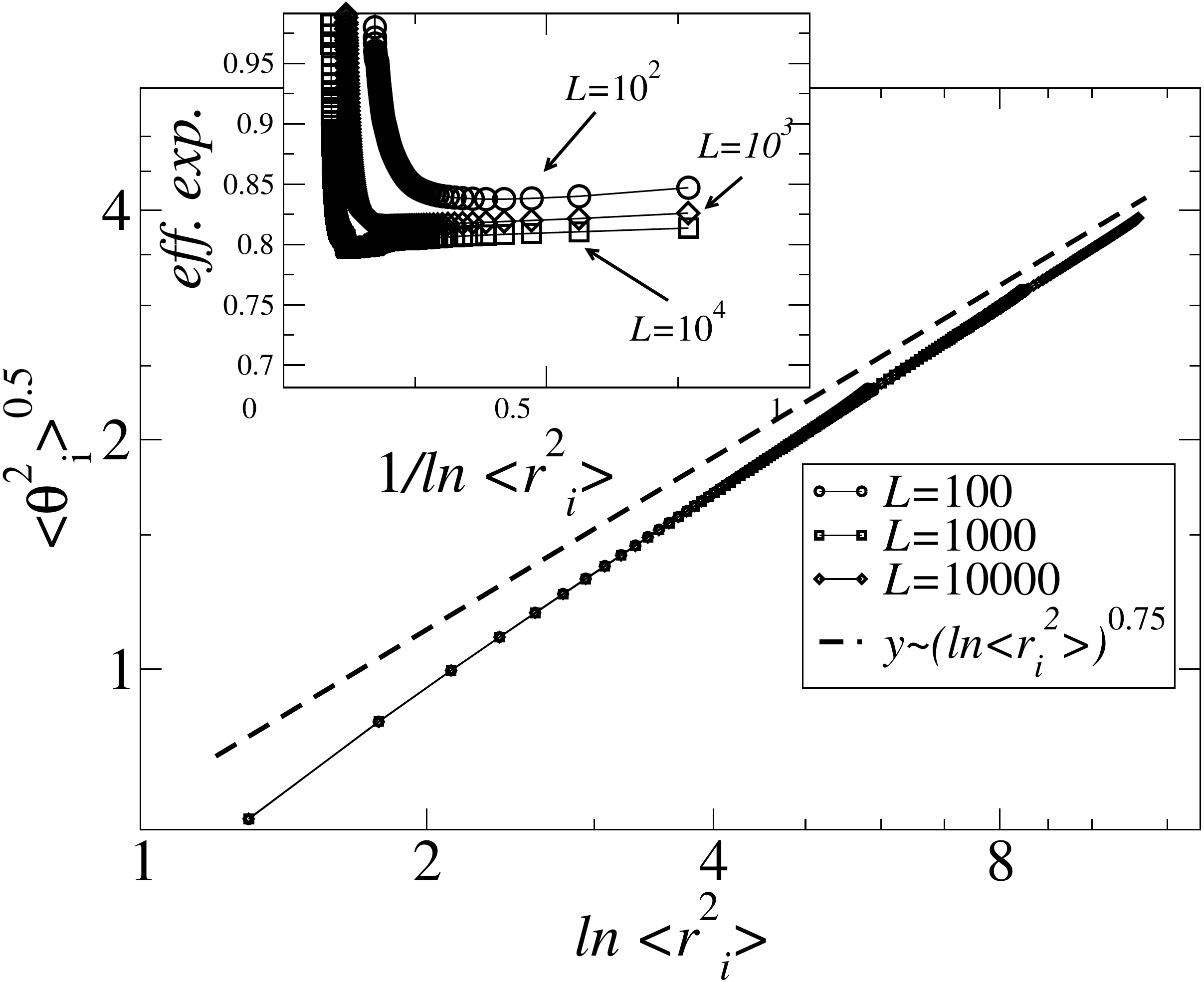}
\caption{Logarithmic plot of $\sqrt{\langle\theta^2\rangle}$
vs. $\ln \langle r^2_i\rangle$ where $\langle r^2_i\rangle$ is the
mean square distance of the monomer $i$ from the monomer attached to
the bar.  Shown are the data for polymers of lengths $L=10^2$,
$10^3$ and $10^4$. Inset: plot of the effective exponents for the
three sizes. Due to strong finite-size effects, the exponent is not converging 
for the different sizes. We can only conclude an upper bound $\alpha<0.8$.}
\label{RvsW}
\end{figure}

Figure~\ref{RvsW} shows a plot of the average winding angle of the $i$-th
monomer as a function of the mean square distance $\langle r_i^2\rangle$
from the monomer $i=1$, which is attached to the bar. Since $r_i \sim
i^{\nu}$ we expect $\sqrt{\langle \theta^2 \rangle} \sim (\ln \langle
r_i^2\rangle)^\alpha$. The best fit of the data in the log-log plot yields
a value $\alpha \approx 0.8$ (dashed line), which is slightly higher than
the previous estimates. The inset of Fig.~\ref{RvsW} shows the plots of
the integrated effective exponents plotted as function of $1/\ln \langle
r_i^2 \rangle$.  As was the case in Fig.~\ref{theta2_vs_i}, we notice a
divergence of the effective exponents for $i \to L$ due to end effects.
We notice also strong finite size effects since corrections to scaling
can arise from both $\langle r_i^2 \rangle$ and $\sqrt{\langle \theta^2
\rangle}$.  An accurate estimate of $\alpha$ is more difficult from
these data.  However, when increasing the polymer length, the effective
exponent decreases (ignoring the end behavior), suggesting as upper
bound $\alpha < 0.8$.

Summarizing, the analysis of Monte Carlo
data yield a consistent value of $\alpha \approx 0.75(1)$, at least for the
first two quantities analyzed.  We note
that the computed value of $\alpha$ is between the two-dimensional random
walk case $\alpha =1$ (Eq.~\ref{disk_abs}) and the two-dimensional
self-avoiding walk case $\alpha = 1/2$ (Eq.~\ref{saw_2d}). It can be
interpreted as follow. Projecting the polymer configuration onto
a plane perpendicular to the bar one obtains
a two-dimensional projection where the walk has some overlaps, but the
three-dimensional self-avoidance constraints reduces these overlaps
compared to those of a full planar random walk.

\begin{figure}[t]
\includegraphics[width=0.9\columnwidth]{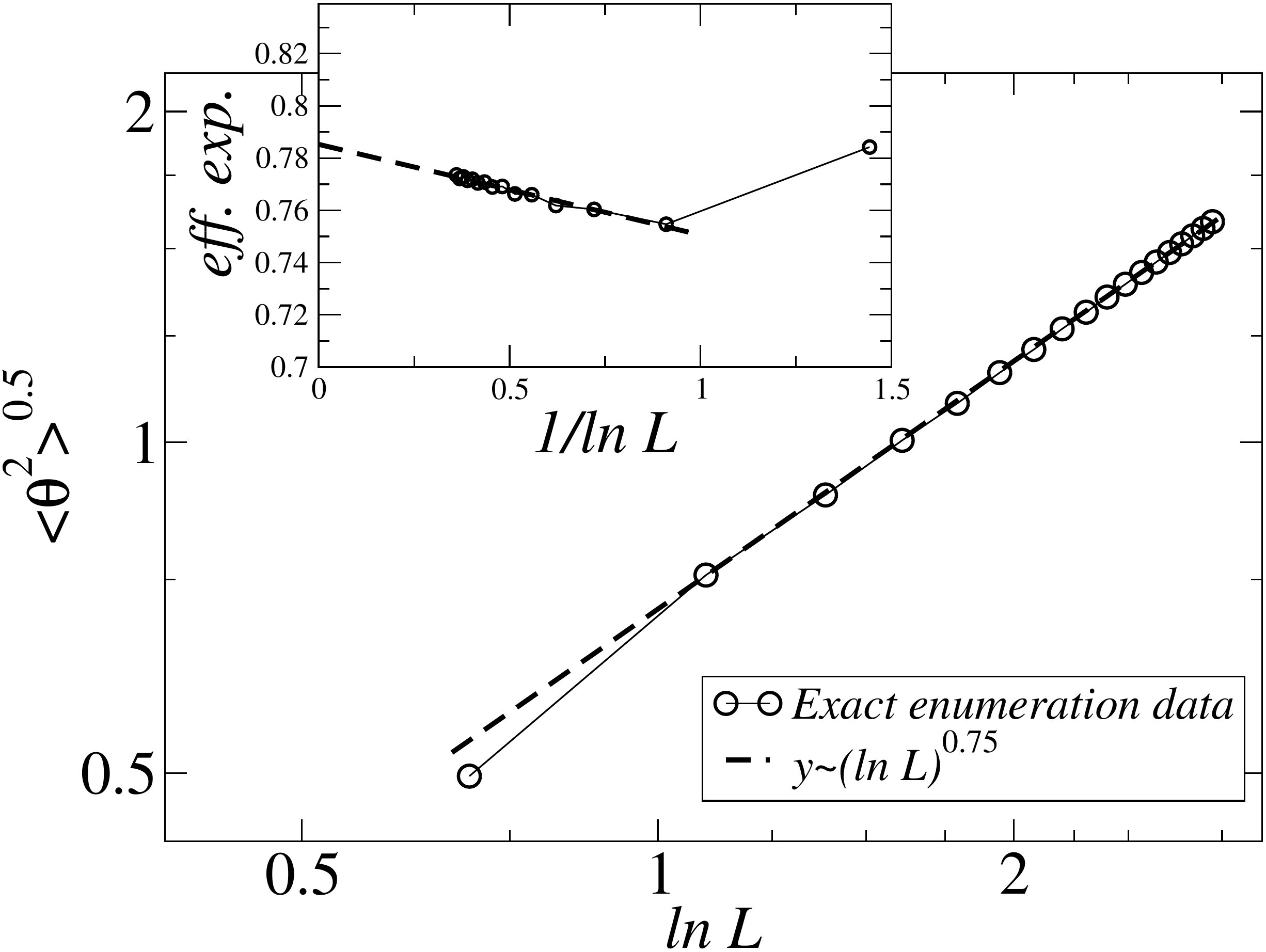}
\caption{Double-logarithmic plot of $\sqrt{\langle\theta^2\rangle}$
vs. $\ln L$ from the exact enumeration data. Inset: plot of the
effective exponent as a function of $1/\ln L$. A linear
extrapolation of the effective exponent using a $1/\ln L$ variable gives
$\alpha\approx0.78$, which is close to the estimate from Monte Carlo data. 
}
\label{ExpExactEn}
\end{figure}

To corroborate these finding we also performed exact enumerations of
a self-avoiding walk winding around a bar on the cubic lattice. As
the number of walks grows exponentially in $L$ the calculations
were restricted to $L \le 20$.  The table~\ref{table1} gives the total
number of walks, the average squared end-to-end distance $\langle
r^2\rangle$, the average squared winding angle $\langle\theta^2\rangle$
and the ratio $\langle\theta^4\rangle/\langle\theta^2\rangle^2$. In
Figure~\ref{ExpExactEn} we show a log-log plot of $\sqrt{\langle \theta^2
\rangle}$ as a function of $\ln L$.  These data are consistent with those
obtained from the Monte Carlo analysis, namely of a power-law behavior
$\sqrt{\langle \theta^2 \rangle} \sim (\ln L)^\alpha$ with $\alpha
\approx 0.75$. The calculation of the effective exponent is plotted as
a function of $1/\ln L$ in the inset of Fig.~\ref{ExpExactEn}.  As seen
from the data the effective exponent is linearly related to $1/\ln L$.
A linear extrapolation for infinite sizes using $1/\ln L$ as scaling
variable gives $\alpha\approx0.78$, which is close to the estimate
from Monte Carlo simulations.
Strong oscillations for odd-even sizes do not allow the use of more refined
extrapolation methods such as the BST algorithm~\cite{henk88}.

\begin{table}[t]
\begin{center}
\caption{Exact enumeration results for a self-avoiding walk on a
cubic lattice, which is attached to a bar. For sizes ranging from
2 to 20 steps.  Measured are the number of walks, their average
squared end-to end distance $\langle r^2\rangle$, the average
squared winding angle $\langle \theta^2\rangle$, and the ratio
$\langle\theta^4\rangle/\langle\theta^2\rangle^2$.
}
\begin{tabular}{|rrrrl|}
\hline
Size &  Num. walks    & $\langle r^2\rangle$ & $\langle\theta^2\rangle$&$\langle\theta^4\rangle/\langle\theta^2\rangle^2$\\
\hline
2    &  20            & 2.400000     &  0.246740 &2.50001\\
3    &  92            & 4.130435     &  0.573089 &2.43716\\
4    &  444           & 5.801802     &  0.801767 &2.69437\\
5    &  2076          & 7.736031     &  1.008331 &2.90674\\
6    &  9860          & 9.619473     &  1.177011 &3.09368\\
7    &  46356         & 11.654241    &  1.338818 &3.18679\\
8    &  219316        & 13.667001    &  1.475501 &3.28308\\
9    &  1031836       & 15.795530    &  1.608453 &3.33777\\
10   &  4871212       & 17.908542    &  1.725029 &3.39814\\
11   &  22917588      & 20.115627    &  1.837856 &3.42916\\
12   &  108046716     & 22.311513    &  1.939638 &3.47006\\
13   &  508228828     & 24.585907    &  2.037864 &3.48878\\
14   &  2393946452    & 26.852009    &  2.128326 &3.51787\\
15   &  11257861180   & 29.185833    &  2.215473 &3.53001\\
16   &  52994270612   & 31.513181    &  2.296973 &3.55151\\
17   &  249151623836  & 33.900291    &  2.375421 &3.55988\\
18   &  1172249039916 & 36.282137    &  2.449638 &3.57625\\
19   &  5510044713020 & 38.717590    &  2.521054 &3.58235\\
20   &  25914234060972& 41.148633    &  2.589254 &3.59503\\
\hline
\end{tabular}
\label{table1}
\end{center}
\end{table}

\begin{figure}[t]
\includegraphics[width=0.9\columnwidth]{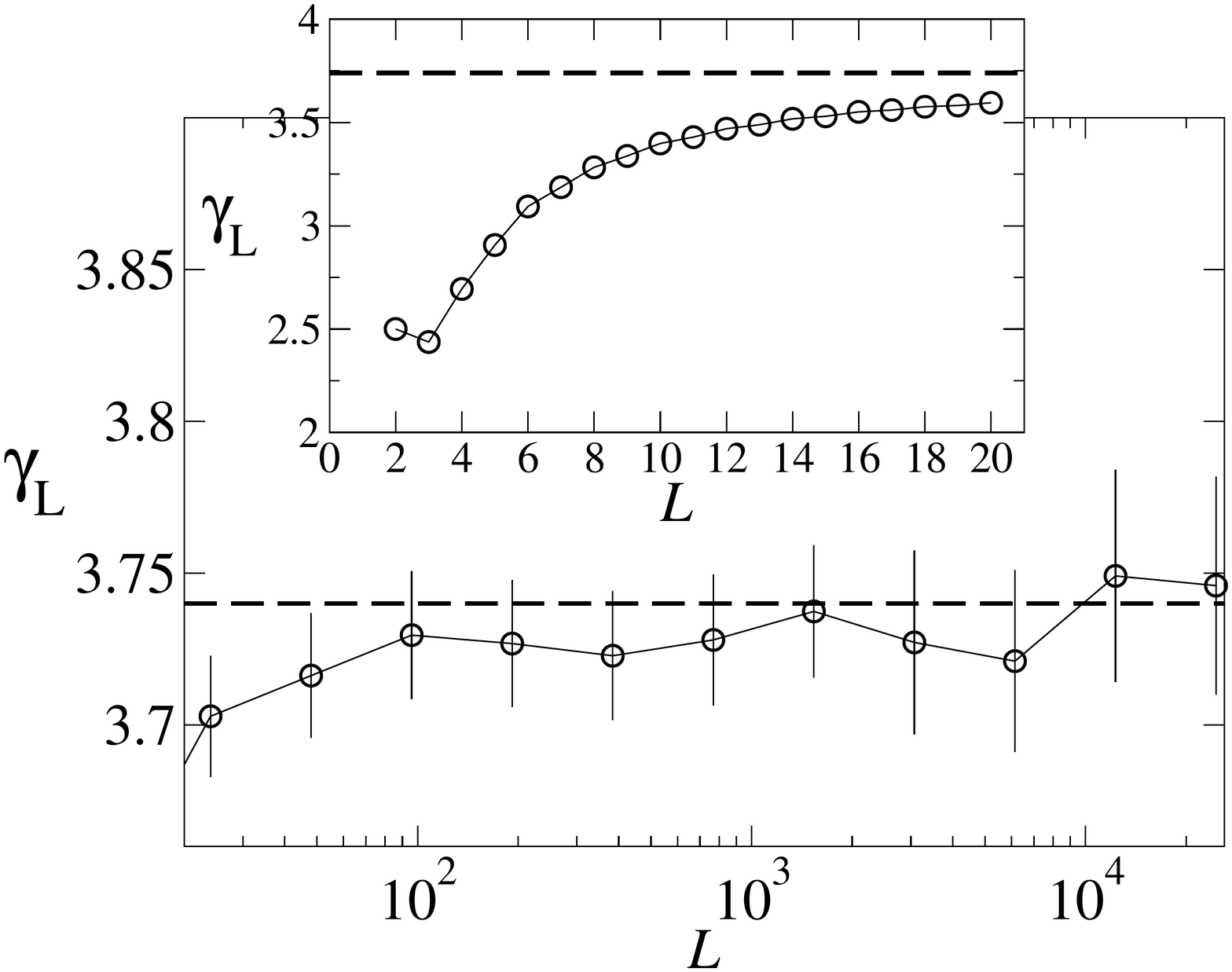}
\caption{Dependence of the ratio $\frac{\langle \theta^4\rangle}{{\langle
\theta^2\rangle}^2}$ with respect to the size. Main graph: we observe
fluctuations around $3.74(5)$ but no shift toward the value $3$ of the
gaussian distribution. For convenience of the reader, the horizontal axis
is logarithmic.  Insert: data from exact enumeration of
the simple cubic lattice.  The dotted line indicates the value obtained
in the main graph.}
\label{ratio}
\end{figure} 

\subsection{The probability distribution function (pdf)}

We consider now the pdf of winding angles $p(\theta,L)$. Equivalently,
this is related to the free energy as a function of winding angle,
using the relation
\begin{equation}
\beta F = -\ln \left( p(\theta,L)/p(0,L)\right).
\end{equation}

A direct evidence of the non-gaussian behavior of this distribution is
provided by Monte Carlo simulations with the same
parameters as the previous subsection.  The main frame of Fig.~\ref{ratio}
shows the value of the ratio
\begin{equation}
\gamma_L = \frac{\langle \theta^4 \rangle}{\langle \theta^2 \rangle^2}
\end{equation}
plotted for differents sizes. We obtain the estimate $\gamma = 3.74(5)$,
which is a value well above the expectation for a gaussian distribution
($\gamma = 3$ for a gaussian pdf).  The exact enumeration data reported
in the last column of Table~\ref{table1} for different polymer lengths
are shown in the inset of Fig.~\ref{ratio}. They display a convergence in good agreement
with the value obtained from Monte Carlo simulation (dashed line).

\begin{figure}[t]
\includegraphics[width=0.9\columnwidth]{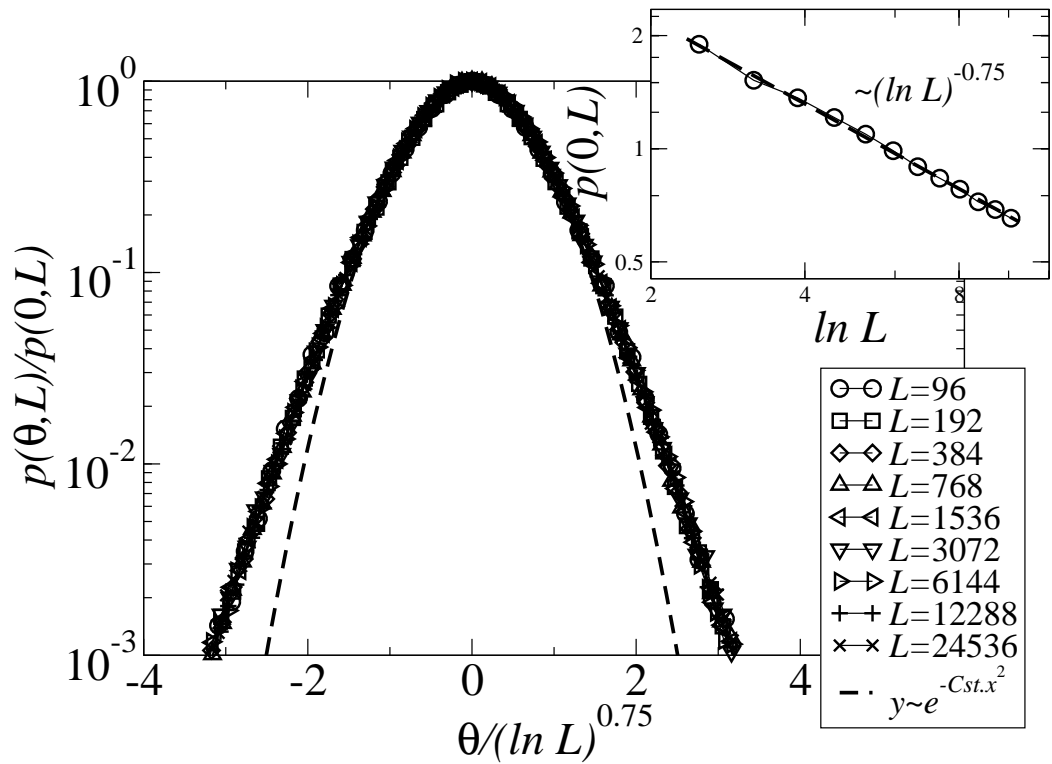}
\caption{Histogram of the winding number of the polymer around the bar.
The probability is normalized and plotted vs. $(\ln L)^{0.75}$.  The data
display a scaling behavior, confirming the power law found above.
The dotted line is a gaussian fit which clearly  does not fit the data
as $\theta$ increases. Insert~: $p(0,L)$ is plotted vs. $\ln L$ and
behaves as $p(0,L)\sim(\ln L)^{-0.75}$.}
\label{theta_beta0}
\end{figure}

The full plot of the winding angle distribution function is shown
in Fig.~\ref{theta_beta0} for Monte Carlo simulations.
 The binning of the winding angles
is done with intervals of $\approx0.5$ rad.  The data for the different
lengths collapse very well onto a single curve when the scaling variable
$x = \theta/(\ln L)^{0.75}$ is used, a value which confirms the exponent
obtained from the analysis of the variance of the winding angle. The inset
of Fig.~\ref{theta_beta0} shows a log-log plot of $p(0,L)$ as a function
of $\ln L$ confirming again the scaling exponent $\alpha\approx 0.75$.
The pdf is thus described by a scaling form of the type
\begin{equation}
p(\theta, L) = \frac{Cst.}{(\ln L)^\alpha} f\left( \frac{\theta}{(\ln L)^\alpha}\right)
\end{equation}
where $f(x)$ is a scaling function. The dashed line in
Fig.~\ref{theta_beta0} shows a parabolic fit of the data. It indicates
that the tail of the distribution decays slower than that of a gaussian
distribution.

\begin{figure}[t]
\includegraphics[width=0.9\columnwidth]{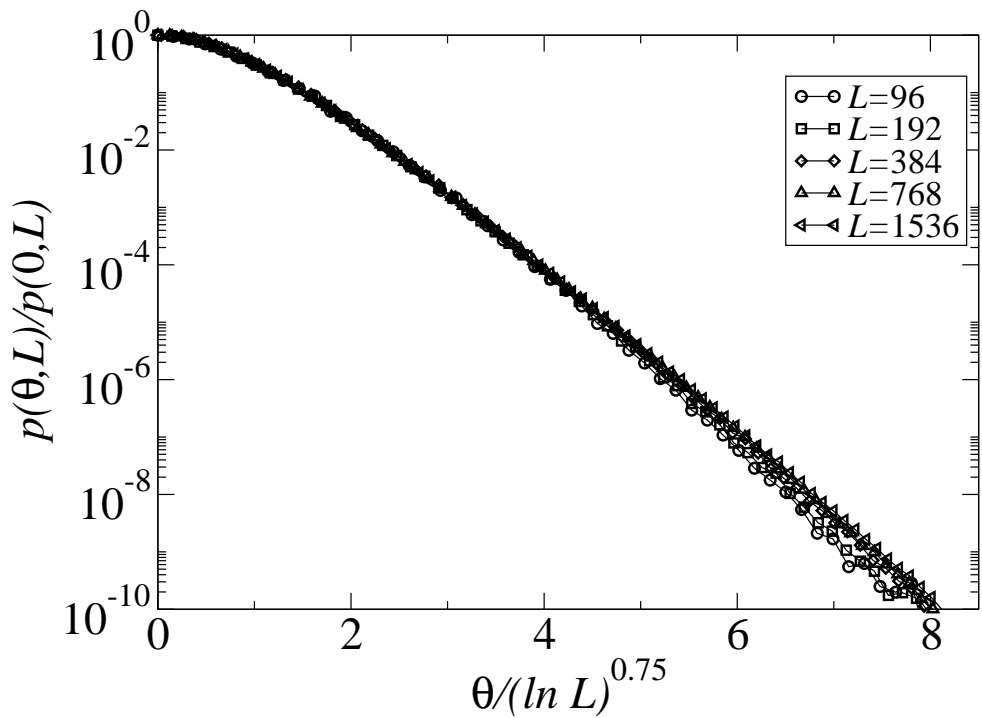}
\caption{Histogram of the winding angle of the polymer around the bar. 
The probability is estimated through biased sampling, such that the region of 
large winding and low probability is visited. The two smallest sizes start to display
finite-size effects for large winding angles. However, the three largest sizes 
show a good collapse with the scaling variable $\theta/(\ln L)^{0.75}$.
}
\label{theta_beta_rew}
\end{figure}

In order to sample the tail of the distribution with a
sufficient statistical accuracy, i.e., beyond the data shown in
Fig.~\ref{theta_beta0}, we have used a reweighting technique. A weight
$\exp(A(\theta_{a}-\theta_{b}))$ was added in the acceptance probability
of each Monte Carlo move.  The angles $\theta_{b}$ and $\theta_{a}$
are respectively the angles before and after the Monte Carlo trial
move and $A$ is a positive constant.  This favors large winding angles
depending on the value of the constant $A$. The distributions obtained
are then multiplied by $\exp(-A\theta)$ to retrieve unbiased results.
The results are shown in Fig.~\ref{theta_beta_rew} for lattices sizes
ranging between $L=96$ and 1536 by steps of a factor of two. The results
are averaged over $10^7$ configurations, i.e., five times more than the
histogram of Fig.~\ref{theta_beta0}, so we restricted ourselves to smaller
sizes compared to those of Fig.~\ref{theta_beta0}.  We have chosen the
value $A=0.45$ for all sizes. For the two smallest sizes, finite-size
effects occur for large winding. However, other than that, the data
display a nice scaling collapse until $p(\theta)/p(0)\approx10^{-10}$.
These results confirm the scaling behavior observed at smaller winding
angles on Fig.\ref{theta_beta0}.

\begin{figure}[ht]
\includegraphics[width=0.9\columnwidth]{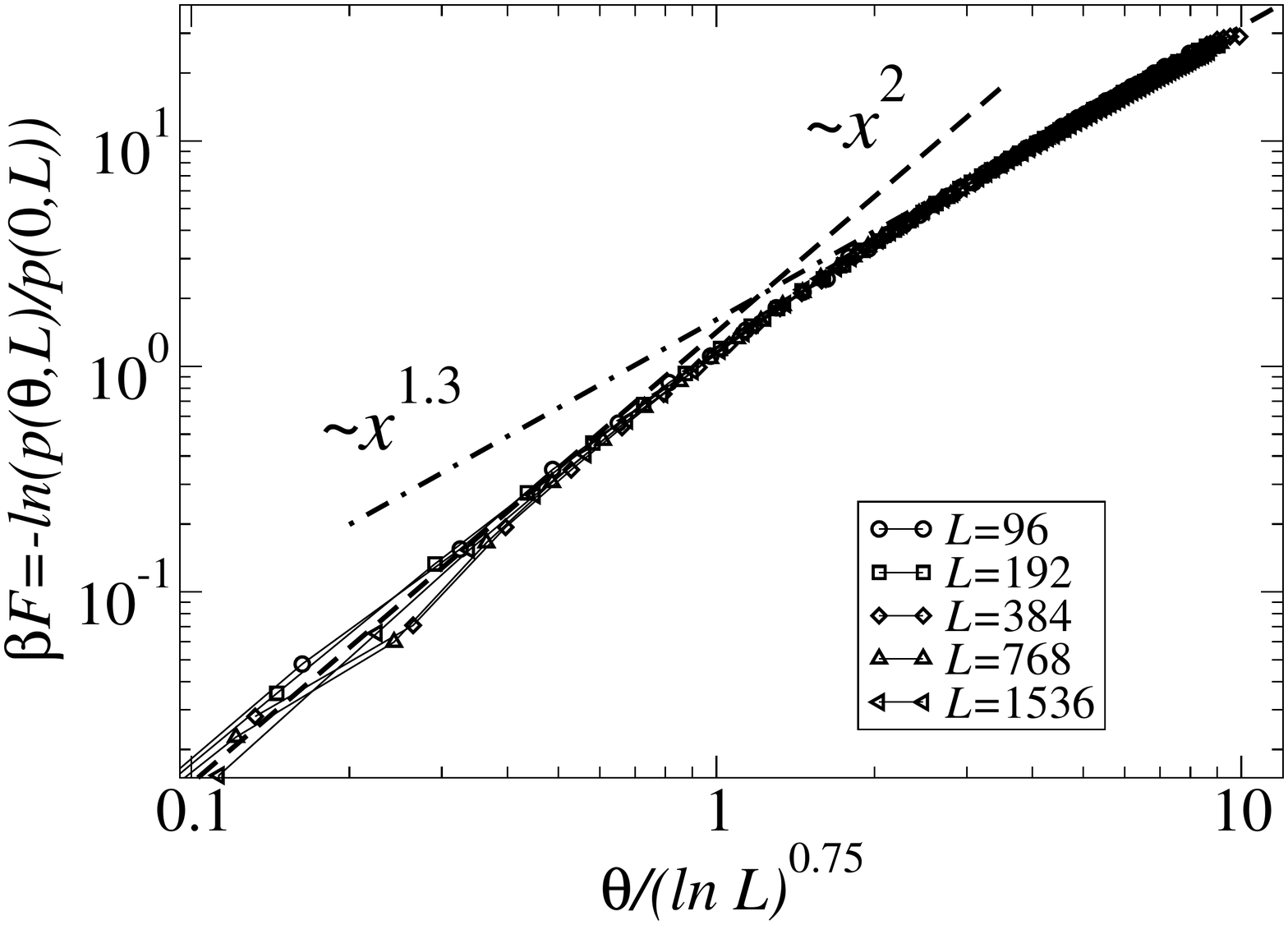}
\caption{Free energy plotted as function of the
scaling variable $x=\theta/(\ln L)^\alpha$ for polymers of various
lengths. This quantity shows a crossover from $\sim x^2$ behavior
at small $x$ (dashed line) towards a different regime at higher winding. The
fit of the data at high winding yields $\sim x^{1.3}$ (dashed-dotted line). 
}
\label{fig_free_ene_rew}
\end{figure}

Figure~\ref{fig_free_ene_rew} shows the same data as in
Fig.~\ref{theta_beta_rew}, this time showing the free energy as a function
of the scaling variable $\theta/(\ln L)^\alpha$ with $\alpha = 0.75$,
in a double-logarithmic plot.  For small winding angles the free energy
increases quadratically in $\theta$, which is the harmonic response
to small winding.  At higher $\theta$'s, the shape of the free-energy
curve changes and follows a different behavior, which corresponds to
the deviation from the gaussian shape observed in the pdf.

Although the range of $\theta$ one can analyze with the reweighting
method described above is somewhat limited, the data suggest a possible
crossover from the gaussian behavior $F \sim x^2$  at small
 $x= \theta/(\ln L)^\alpha$ i.e. small $\theta$, to a different power-law.
 A power-law fit in the region of high winding numbers yields a behavior
 $F \sim x^{1.3}$.  This behavior is reminiscent of the scaling
of the free energy for the gaussian linking number of two tethered
rings~\cite{mark09}. The linking number is a topological invariant
and indicates the degree of entanglement of two polymer rings. This
quantity roughly represents the number of times that each rings winds
around the other, and can thus be considered an analogous of the winding
angle $\theta$ of a polymer wound around a bar.

For random walks, the free energy as a function of the linking number
$Ca$ was found to scale as~\cite{mark09} $F \sim Ca^2$ for small $Ca$
and as $F \sim Ca^{4/3}$ at stronger $Ca$, from simulation data and
analytical arguments. It is however not yet clear whether this analogy
holds for the winding angle distribution, as for self-avoiding tethered
rings the free energy is linear in the linking number $F \sim Ca$,
suggesting phase separation.

We propose the following arguments to justify the non-gaussianity of the curves. 
As noted before, the maximal winding number that can be achieved with a
polymer of length $L$ is $\theta_{max}=2\pi L/6$. An upper limit to the
range in free energy is given by $\Delta F \le \beta^{-1} \log(Z_L)$
where $Z_L$ is the total number of states accessible to the polymer.
Since the set of random-walk configurations is a superset of the set of
self-avoiding-walk configurations wrapped around a bar, it follows that
$Z_L < \mu^L$ with $\mu=6$ for a cubic, and $\mu=12$ for a fcc lattice,
and thus it follows that $\log(Z_L)\le \mu L$.  Consequently, the free
energy of the the maximally wound state can at most increase linearly with
$L$, and certainly not quadratically. Hence, the existence of a maximally
wound state with $\theta \sim L$ together with the maximal range in free
energy of $\sim L/\beta$ already provides proof of nongaussian behavior of
the pdf at large winding.  (Note that for two-dimensional self-avoiding
walks, the same argument does not exclude a gaussian distribution,
as the maximal winding angle is then $\theta_{max}\sim\sqrt L$).

\subsection{The polymer shape}

We consider next the equilibrium shape of the polymer wound around the bar
at some fixed winding angle $\theta$.  We use cylindrical coordinates
where the bar is the reference axis and the origin is the point of
attachment of the polymer. $R_i$ is the radial coordinate where $i$
labels the monomers, starting from the monomer attached to the bar.  $z_i$
is the coordinate in the direction parallel to the bar, and $\omega_i$
is the winding angle of the monomer $i$ along the polymer. The total
winding angle is given by $\omega_L=\theta$.

In order to sample configurations with high winding we have performed
reweighted simulations as explained in the previous subsection, using
the parameter $A=0.45$.  In this way a broad range of values of $\theta$ were
generated in the course of the simulation. Each time a configuration with
a winding angle equal to $\theta_n = n \pi/2$ with $0 \leq n \leq 14$
was generated the cylindrical coordinates  $(R_i, \omega_i, z_i)$
were stored. In this way the averaged shape was computed
for some selected winding angles $\theta_n$ (in practice we sampled
over all configurations with winding in the interval $[\theta_n - 0.01,
\theta_n + 0.01]$).

\begin{figure}[t]
\includegraphics[width=0.85\columnwidth]{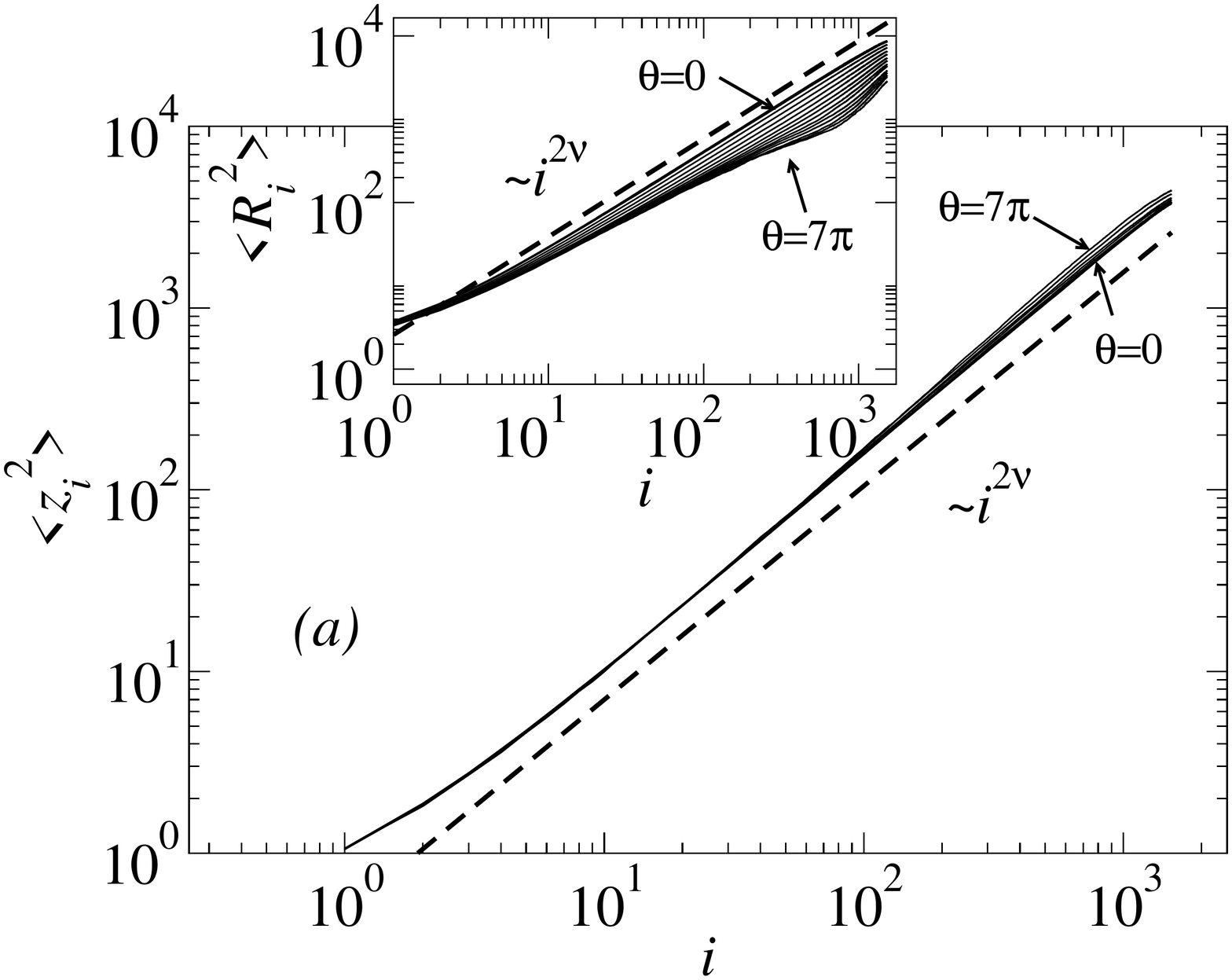}

\includegraphics[width=0.85\columnwidth]{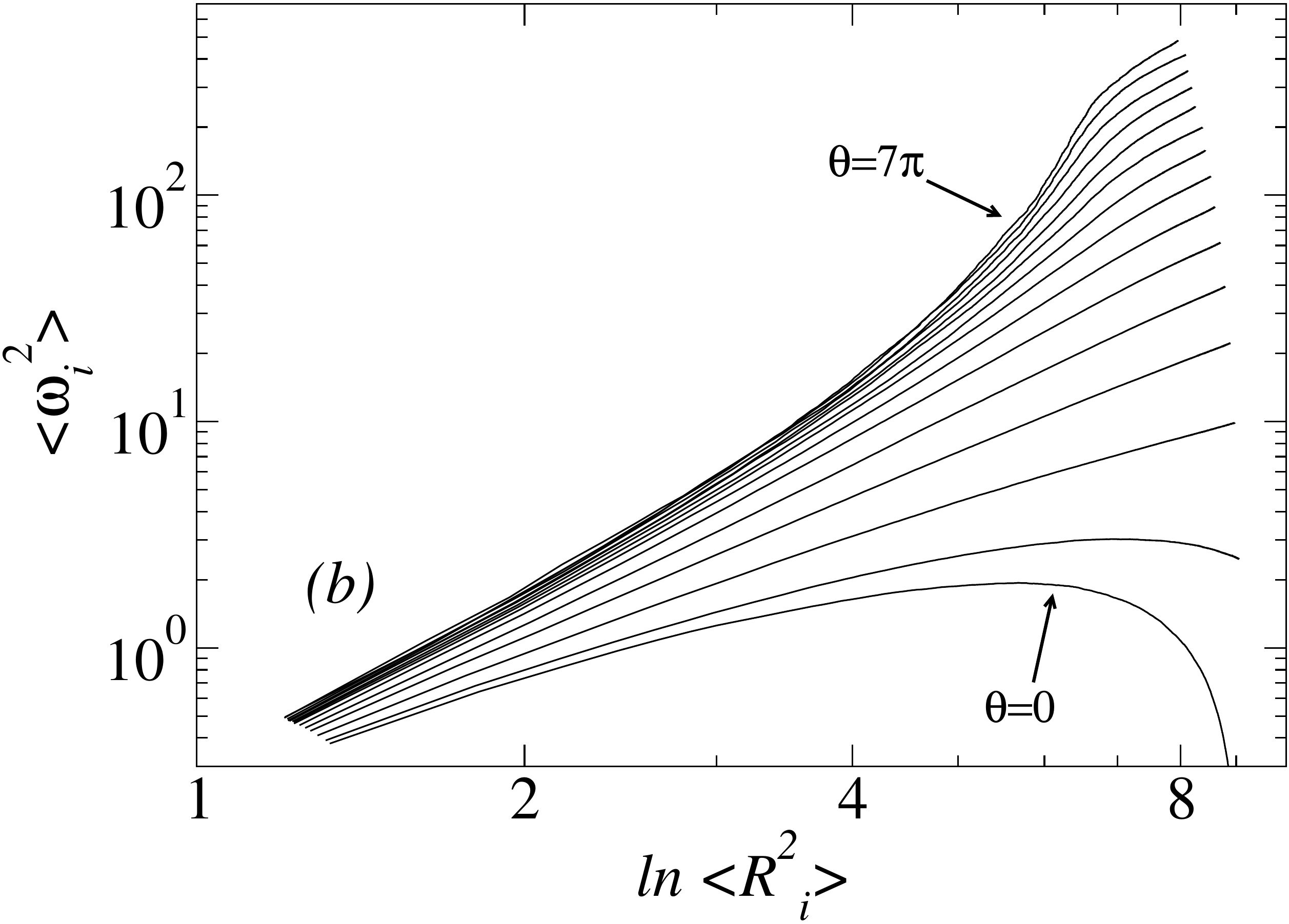}
\caption{Configuration of a polymer of size $L=1536$ for different
winding angles $\theta$.  Top:  the coordinate $z_i$ (main graph)
 and $R_i^2$ (inset) versus the monomer index $i$.
 Bottom: the two relevant coordinates of the system: $\omega^2_i$ versus $\ln R^2_i$.
For intermediate values of $\theta$, they display a power law relation.}
\label{fig_shape}
\end{figure}

The range of values of $\theta$ analyzed spans completely the region shown
in Fig.\ref{theta_beta_rew}, i.e., the region containing the equilibrium
configurations with a significant probability.  Figure~\ref{fig_shape} (top)
plots the behavior of $z_i^2$ vs $i$ (main graph) and of $R_i$ vs $i$
(insert) for different total winding angles $\theta_n$.  For all values of
$\theta_n$ investigated, the data follow rather well the self-avoiding
walk scaling $z_i^2 \sim i^{2\nu}$, where $\nu =0.588$ is the Flory
exponent~\cite{vand98}. In addition, the data corresponding to different
winding angles $\theta$ overlap, showing that the distribution of the
monomers in the direction parallel to the bar is weakly influenced by
the total winding angle.  The inset of Fig.~\ref{fig_shape} (top) shows the
behavior of $R_i^2$ vs. $i$. In this case there is a stronger dependence
on the winding angle. For a given $i$, the distance from the bar $R_i$
decreases at higher winding angles, as expected. For the smallest winding
(top curve), the data follow quite well the scaling $R_i^2 \sim i^{2\nu}$.
For higher winding angles $R_i$ keeps increasing monotonically with $i$,
but the data show a kink close to the polymer end.

Figure~\ref{fig_shape} (bottom) plots $\omega^2$ vs. $\ln R^2$ for different
total winding angles $\theta$, obtained as discussed above. The data
are plotted on a log-log scale. The lowest curve corresponds to the
smallest winding angle analyzed ($\theta=0$). We note that the behavior
is non-monotonic, implying that inner monomers can have a higher winding,
compared to the end monomer.  For higher total winding angles, the curves
are monotonic. The values of $\theta$ for which the crossover between
non-monotonic to monotonic scaling occurs, corresponds approximately
to the crossover between quadratic to non-quadratic response in the
free energy discussed in the previous subsection.  A straight line
in the plot of Fig.~\ref{fig_shape} (bottom) would correspond to a shape 
described by a stretched exponential
\begin{equation}
R(\omega) \sim e^{\Gamma \omega^\gamma}
\end{equation}
Although for a few intermediate winding angles the lines appear to be
rather straight in the graphs, it is difficult to capture the shape
with a simple analytical form. However, these shapes could be compared
with those obtained during unwinding dynamics~\cite{baie10a}.

\section{Conclusion and future work}

In this paper we have investigated the equilibrium behavior of a polymer
wound around an infinitely long bar. The polymer is self-avoiding and it
has excluded volume interactions with the bar as well.  The main result
of the paper is that the pdf is described by a scaling variable of the
type $\theta/(\ln L)^\alpha$, where $\alpha \approx 0.75$.  The fact that
scaling involves the logarithm of the polymer length is not surprising,
as this is also found in the case of two-dimensional self-avoiding and random
walks. The exponent $\alpha$ in those two cases is however different with
$\alpha =1$ for planar random walks~\cite{rudn87} and $\alpha = 1/2$
for self-avoiding walks~\cite{dupl88c}. The case of a three-dimensional
polymer appears to be intermediate between the two.  The presence of a
logarithm is responsible for slow asymptotic convergence and some care has
to be taken in this case. Our analysis, however, involves very long polymers
with $L \approx 25\,000$, and thanks to a reweighting technique, explores
high winding numbers, i.e., low-probability regions of the distribution. In
addition several different quantities have been analyzed and all confirm
$\alpha \approx 0.75$. The same holds for exact enumeration results.

The pdf itself appears to deviate from a gaussian behavior. 
The ratio $\gamma=\langle\theta^4\rangle/\langle\theta^2\rangle^2=3.74(5)$
 differs from the gaussian value $\gamma=3$. This is best
characterized by the scaling of the free energy which crosses over
from $F \sim x^2$ at small $x=\theta/(\ln L)^\alpha$, towards a behavior
characterized by a different power-law $F \sim x^{1.3}$, although in a small
range of values of $x$. 

More analytic insights in these problems are lacking at the moment. These
are restricted to a first-order $\varepsilon$-expansion around four
dimensions~\cite{rudn88}. These results suggest a scaling behavior with
$\alpha = 1/2$ also in three dimensions which is at odds with our present
numerical findings.

Our future work will proceed along various lines. First, to connect
our work more directly with DNA melting, we intend to introduce an
attractive interaction between the polymer and the bar, mimicking the
hybridization between the two DNA-strands.  This interaction will
initially be homogeneous, but later might have a random component
to capture the difference between the binding strengths of AT and
CG bonds.  How such interactions influence the equilibirum statistics
of winding, is an open issue. Secondly, we want to combine the results
presented here with earlier work involving some of us on the unwinding
dynamics~\cite{baie10a}, which is still an open issue.

\section*{References}


\end{document}